# Gradual Soundness: Lessons from Static Python

Kuang-Chen Lu[a], Ben Greenman[a], Carl Meyer[b], Dino Viehland[b], Aniket Panse[b], and Shriram Krishnamurthi[a]

a Brown University, Providence, RI, USA

b Meta, Menlo Park, CA, USA

**Abstract**

**Context** Gradually-typed languages allow typed and untyped code to interoperate, but typically come with significant drawbacks. In some languages, the types are unreliable; in others, communication across type boundaries can be extremely expensive; and still others allow only limited forms of interoperability. The research community is actively seeking a sound, fast, and expressive approach to gradual typing.

**Inquiry** This paper describes Static Python, a language developed by engineers at Instagram that has proven itself sound, fast, and reasonably expressive in production. Static Python's approach to gradual types is essentially a programmer-tunable combination of the *concrete* and *transient* approaches from the literature. Concrete types provide full soundness and low performance overhead, but impose nonlocal constraints. Transient types are sound in a shallow sense and easier to use; they help to bridge the gap between untyped code and typed concrete code.

**Approach** We evaluate the language in its current state and develop a model that captures the essence of its approach to gradual types. We draw upon personal communication, bug reports, and the Static Python regression test suite to develop this model.

**Knowledge** Our main finding is that the *gradual soundness* that arises from a mix of concrete and transient types is an effective way to lower the maintenance cost of the concrete approach. We also find that method-based JIT technology can eliminate the costs of the transient approach. On a more technical level, this paper describes two contributions: a model of Static Python and a performance evaluation of Static Python. The process of formalization found several errors in the implementation, including fatal errors.

**Grounding** Our model of Static Python is implemented in PLT Redex and tested using property-based soundness tests and 265 tests from the Static Python regression suite. This paper includes a small core of the model to convey the main ideas of the Static Python approach and its soundness. Our performance claims are based on production experience in the Instagram web server. Migrations to Static Python in the server have caused a 3.7 % increase in requests handled per second at maximum CPU load.

**Importance** Static Python is the first sound gradual language whose piece-meal application to a realistic codebase has consistently improved performance. Other language designers may wish to replicate its approach, especially those who currently maintain unsound gradual languages and are seeking a path to soundness.

## 1 Introduction

Gradual typing is an enticing solution to the debate between static and dynamic typing. The premise is simple: let programmers introduce types in part of a codebase while leaving the rest untyped. Run-time casts and checks can enforce the assumptions that typed code makes about untyped components, thereby ensuring that the types are sound no matter how untyped code behaves.

Unfortunately, the high run-time cost of sound types has split the gradual typing community. Industry teams have developed innovative *optional* type systems that accommodate untyped designs, but are unsound [2, 7, 45, 63]. These systems intentionally check nothing at run-time when untyped values enter typed code. Academic teams have primarily focused on the theory of sound gradual types, formulating correctness properties and studying ever-more-descriptive type systems (e.g., [6, 34, 41, 51]). A few academics have studied the cost of run-time checks in detail [23, 55] and proposed implementation methods [11, 30], compiler technology [1, 65], and even weakened semantics [21, 25, 66], but these efforts have not yet decisively closed the performance gap. The most promising attempt is the *concrete* semantics for gradual types [8, 39, 67], but it imposes nonlocal constraints on untyped code. In particular, concrete-typed client code is incompatible with values created by untyped code.

In short, academic researchers are working to close the performance gap without overly restricting the promise of gradual typing. Industry researchers are sidestepping the problem with unsound types — for the most part.

This paper reports an exception to the rule among industry-made gradual type systems. The Static Python team at Instagram has developed a sound type system for a subset of Python along with a runtime system that uses soundness to drive optimizations. The language is staged to let programmers choose between easy migrations and full-strength optimizations, a design that we call *gradual soundness*. To a first approximation, there are three main stages:

1. A full language of *shallow types* that describe idiomatic Python code at a coarse granularity. Adapting an untyped module to use shallow types requires at most a few local code changes.
2. A second level of *concrete types* describes generic data structures that check the types of their elements. If programmers modify their code to build concrete structures instead of Python ones (a potentially nonlocal change), then Static Python can perform deeper optimizations.
3. A third set of progressive types enable further optimizations. In particular, *primitive types* allow for unboxed arithmetic.

Over the past year, Instagram has been applying gradual soundness to its primary web server monolith. Although only a handful of modules use Static Python types (hundreds among thousands), and only a few critical modules rely on concrete types and primitive types (dozens), these migrations have resulted in a 3.7 % increase in production requests handled per second for servers running at maximum CPU load. These results are very positive: Type-directed optimizations outweigh the cost of enforcing soundness. Consequently, we believe the details of the Static Python approach are of interest to the gradual typing community at large.

**Contributions** This paper makes two contributions:

- *Evidence*. We present evidence that Static Python improves performance with few code changes / refactorings. The Instagram web server has become significantly more efficient thanks to a gradual application of Static Python to high-profile modules. Because the server code is closed-source, we additionally present data for four public microbenchmarks (Section 6.4). Using only shallow types, Static Python runs slightly faster than Python despite the costs of enforcing soundness (avg. 20 %). With fine-tuned types, Static Python consistently outperforms the Python baseline (avg. 70 %).
- *Mechanization*. To validate the soundness of Static Python, we mechanized a core language in PLT Redex and ran both property-based soundness tests and over 250 tests adapted from the Static Python regression suite. The modeling effort revealed 21 significant issues in Static Python, five of which were soundness bugs.

**Outline** This paper begins with an informal description of Static Python in two parts. First, we present a user-oriented summary of the language (Section 2). Second, we present the key ingredients of the runtime system, Cinder, that supports Static Python (Section 3). Section 4 uses a small formal model to introduce our Redex mechanization and to demonstrate that Static Python is based on a sound core. Section 5 notes important aspects of Static Python that are not covered by the model. Section 6 evaluates Static Python; it reports our experience with the language in production and on public microbenchmarks. The paper concludes with related work (Section 7) and a final summary (Section 8).

**Significance** We have written this paper with two audiences in mind. First, we want to encourage system-builders to reproduce the Static Python language design. In particular, the maintainers of optionally-typed languages may wish to focus on shallow types and JIT compilation as a first step toward sound and optimized gradual types. Second, we want to entice researchers. The Static Python type system has many noteworthy restrictions. For example, functions are supported only by shallow types, and method overrides that use the dynamic type are more constrained than overrides in untyped code. Some of these restrictions might be lifted by future research. Others might be useful to adopt in new language designs.

## 2 A Tour of Static Python

Static Python is part of a large codebase that includes a type system, a tailored bytecode, and a method-based JIT compiler. In essence, Static Python is the type system for an entirely new language. The interface that it offers to programmers, however, replicates the standard Python experience. Static Python runs Python 3.8

```
def f(x):
  return x["A"]

f({"A": 1})
```

→

```
from typing import Dict
def f(x: Dict[str, int]):
  return x["A"]

f({"A": 1})
```

→

```
from __static__ import CheckedDict
def f(x: CheckedDict[str, int]):
  return x["A"]

f(CheckedDict[str, int]({"A": 1}))
```

**Figure 1** A first Static Python program and two migrations

programs with minimal changes to their behavior (Section 2.6), and it compiles code on-the-fly to support existing IDEs and developer tools. The advanced features of the Static Python type system are offered on an opt-in basis (*gradual soundness*, Section 2.7) and arranged so that programmers can begin adding types one module at a time.

Figure 1 presents a first example program and two modified versions that utilize Static Python types. The basic program defines a function f and calls it with a dictionary value. Static Python can run this program as-is and even JIT-compile the function. The other two versions use dictionary types:

- The Dict type describes normal Python dictionaries in a *shallow* sense. At compile-time, Static Python uses this type to find basic logical errors. At run-time, Static Python enforces the type by checking that all inputs to f are dictionary values. These checks enable additional optimizations within the function body.
- The CheckedDict type describes a *concrete* dictionary data structure provided by Static Python. Unlike Python dictionaries, these checked dictionaries are guaranteed to contain well-typed keys and values even if they escape to untyped code. As the body of f illustrates, the syntax for using a checked dict is standard. Creating a checked dict, however, requires a type-parameterized constructor call.

These typed versions demonstrate the multi-level nature of Static Python. At one level, there are types that describe standard Python values that can be added to a program with little-to-no code changes. At the next level, Static Python offers special-purpose types with stronger guarantees that impose nonlocal maintenance costs, such as requiring edits to constructor calls.

## 2.1 Type System Context and Design Goals

The Static Python type system is a unique synthesis of ideas from the gradual typing literature and prior work on types for Python. Its syntax for types is based on the definitions in PEP 484 [63]. Its static semantics for types is inspired by optional type checkers; in particular, Pyre [37] and mypy [57]. And its strategy for run-time checks is adapted from Nom [39, 40], a research language with compelling performance results. Static Python's novelty comes from the following engineering constraints:

- Performance is the bottom line. At the end of the day, Static Python needs to make code run faster in production.
- Critical code often relies on first-order functions and objects, at least in the Instagram web server. Precise types for higher-order functions, for first-class classes, and even for locally-defined functions and classes are thus a low priority; instead, the dynamic type serves as a coarse approximation.

- The codebase in which Static Python is being applied makes heavy use of PEP 484 type annotations and is regularly analyzed by Pyre. Static Python is therefore compatible with the PEP 484 syntax to reduce the adoption burden.
- A module-level granularity is acceptable. Once Static Python is enabled for a module, it compiles all code in that module including expressions that have the dynamic type (Section 2.2). If this behavior is problematic for certain expressions, they must move to an untyped Python module.

In general, the types that Static Python implements all enable significant optimizations and can be validated quickly with casts. Types that do not meet these criteria are deferred to the mature optional type checkers that already exist for Python. In fact, Static Python has no short-term plans to support all PEP 484 types.

The core supported types describe basic data (integers, booleans, strings), data structures (lists, dicts, promises), and nominal classes. Union types are tracked statically; for example, if x is an integer and y is a string, then the expression x or y has a union type. Unions also narrow down via type tests as a kind of occurrence typing [27, 61]. Union types are not, however, generally supported at run-time (Section 4.2). The only exception is binary unions with the Python None type (Union[S, None], or Optional[S]); these unions are enforced with run-time checks.

Three other unsupported types bear special mention: first-class class and object types (Section 2.4); first-class function (callable) types; and recursive types. Pyre and mypy have some support for the first two kinds, but no support for recursive types.

### 2.2 Type Dynamic

Static Python includes a dynamic type that allows untyped expressions within a statically-typed context. Whenever an expression or variable lacks a type annotation, Static Python uses the dynamic type as a default and skips most static checks. The overall goal of this design is to let programmers add simple annotations to part of a Python module while leaving the rest untyped.

In addition to unannotated positions, the dynamic type is also the default for PEP 484 types that Static Python does not yet understand. For example, if existing code uses the type Set[Int] for sets of integers, then Static Python replaces it with dynamic. Thus, Static Python focuses on types that it can soundly optimize without forcing programmers to remove types that other systems (e.g., mypy) can check.

The behavior of the Static Python dynamic type is subtle. Dynamic-typed code faces more restrictions than untyped Python code. Because of the restrictions, replacing part of a type with dynamic can lead to either a static error or a run-time error. Figure 2 presents two examples, one for each kind of failure:

- Figure 2a presents a fully-typed class and a partially-typed subclass. The subclass definition raises a compile-time error because it attempts to override the typed m method to return the dynamic type.

```python
class A:
  def m(self) -> int:
    return 0

class B(A):
  def m(self):
    # Error: dynamic cannot override int
    return 0
```

**(a)** SGG Violation: removing a type in the subclass B raises a static error

```python
from __static__ import CheckedDict

def f(x: CheckedDict[str, dynamic]):
  return x["A"]

d = CheckedDict[str, int]({"A": 1})
f(d)
# Error: f expected CheckedDict[str, dynamic]
```

**(b)** DGG Violation: removing part of the type for the parameter x raises a dynamic error

**Figure 2** Static Python provides neither the static (SGG) nor the dynamic (DGG) gradual guarantees (assuming a standard type-precision relation)

- Figure 2b sends a checked dictionary with integer values to a function that expects dictionaries with dynamic values. At runtime, the function rejects this argument because the two types are not an exact match.

These extra restrictions make the Static Python dynamic type different than the flexible dynamic type provided by languages that satisfy the gradual guarantees [54]. But they also enable efficient run-time checks for generics and type-directed optimizations for method calls. Part of the reason Static Python weighs these benefits more heavily than the gradual guarantees is that programmers have to opt-in to a feature in order to risk the errors. The static error above comes only after enabling the Static Python compiler on both a parent class and its subclass. The dynamic error comes only after adopting a concrete type for checked dictionaries. In general, Static Python is less interested in guarantees about *removing an arbitrary annotation* and more interested in making sure that *an untyped module compiles with minimal code changes*.

### 2.3 Concrete Types and Shallow Types

Every Static Python type is paired with an efficient cast operation that runs in constant time regardless of how large an incoming value is. Unlike structural gradual type systems [23], no cast traverses an incoming value and no cast creates a wrapper to monitor future behaviors. Some casts for generic types are, however, incomplete. We call the types with incomplete casts *shallow* types and the types with full casts *concrete* types, following the StrongScript authors [49].

First off, casts for non-generic nominal types are straightforward. Every type has a name and every value has a tag that corresponds to the name. In the worst case, a cast must examine the parents of a class to find a match, but these traversals are bounded in length and typically short.

Generic types are more difficult to enforce because a tag by itself does not describe their contents. For example, suppose that a function expects inputs that match the type `Dict[int,int]`; that is, dictionaries with integer keys and integer values. A constant-time check for Python dictionaries is not enough to protect the function against dicts with ill-typed elements. The Static Python solution is to provide a second kind of dictionary that ensures the validity of its elements. These different values have different types:

1. The *shallow* type Dict gets enforced with a tag check for dict values that ignores elements. Consequently, a parameterized surface type such as Dict$[K,V]$ does not promise anything about keys and values.
2. The *concrete* type CheckedDict$[K,V]$ gets enforced with a check for concrete dicts that have exactly the same key and value types. To implement such checks efficiently, Static Python stores a run-time registry of instantiated concrete types. Every concrete dict carries a pointer into this registry to enable constant-time type validation, even for nested concrete types.

**Table 1** Costs of using shallow and concrete dictionaries in typed (T) and untyped (U) contexts relative to the number (N) of dict elements. A dash (—) means zero cost.

| | Cast | T-Read | T-Write | U-Read | U-Write |
|---|---|---|---|---|---|
| Dict | $O(1)$ | $O(1)$ | — | — | — |
| CheckedDict$[K,V]$ | $O(1)$ | — | — | — | $O(1)$ |

These options keep run-time costs low and let programmers choose between concrete and shallow types. This choice is one aspect of gradual soundness in Static Python.

Often, the shallow types are more attractive because concrete types impose nonlocal changes. If one type changes to concrete, then several other changes may need to happen: all values that reach this type must be initialized as concrete, all typed clients of these values must expect concrete values, all values that reach those clients must be concrete, and so on and so forth.

The upside of concrete types is that they enforce stronger type constraints. These constraints can catch bugs and always lead to faster typed code. Faster performance is not guaranteed in general, though, because concrete types must check all writes from untyped contexts. For comparison, Table 1 presents the worst-case costs of shallow and concrete dict types. Shallow types pay a constant-time cost for casts and whenever a typed context reads from a shallow value. Concrete types pay nothing for reads, but incur constant-time costs for casts and for writes from untyped code. If untyped code frequently writes to a value (or equivalently, if untyped code initializes a huge value), then shallow types may run faster.

Not all generic Static Python types have shallow and concrete versions at present, though they are coming soon. In particular, user-defined classes do not have concrete support. The shallow check for a user-defined class does, however, guarantee the types of all non-generic fields and methods.

### 2.4 Gradual Class Hierarchies

One important feature of Static Python is that class hierarchies can mix typed and untyped code. An untyped class can inherit from a typed one and vice-versa, letting programmers add types to a single class independently of its ancestors and children.

Gradual class hierarchies are rarely studied in the literature, especially for a language where truly untyped classes can enter the mix. Thorn [67] and SafeTS [46], for example, provide separate hierarchies for untyped and (gradually) typed classes. The closest related work, for Nom [39], implements a flexible design inspired by the gradual guarantees. In particular, Nom subclasses can override any part of a signature with the dynamic type. Static Python implements a simpler design that restricts types and inheritance:

1. To benefit from types, classes must be declared at the module top level and may have at most one parent. Nested class declarations, first-class classes, and classes with multiple parents default to un-optimized Python behavior (Section 5).
2. Methods cannot be overloaded. This restriction comes from Python.
3. Non-final methods and fields may be overridden in arbitrary ways by untyped code. In typed code, however, overrides cannot use less-precise types. For example, a method that returns a number cannot be overridden by a method that returns the dynamic type; Figure 2a shows the static error that arises from such an override.
4. Typed fields cannot be overridden by descriptors or properties (`@property`).

With this context in mind, Static Python keeps track of whether each class is typed or untyped. Typed classes can assume that if a method has a precise (non-dynamic) type, then all typed overrides are subtypes of this type. This property enables optimized dispatch from typed callers to typed methods and optimized field reads. Other method and field accesses typically incur one run-time check. An extra step arises when an untyped method overrides a typed method. In this case, Static Python creates a wrapper around the overriding method to check that it computes type-correct results. The wrappers are handled efficiently by a tailored implementation of vtables (Section 3).

### 2.5 Progressive Primitive Types

For performance-critical applications, Static Python includes a set of primitive types that describe booleans and sized numbers. Example types include `int64`, `uint64`, `double`. There are also two special datatypes, `Array` and `Vector`, that store primitives. The purpose of these types is to allow unboxed values and arithmetic at runtime. Consequently, their static semantics is a *progressive* [44] refinement over the semantics of basic Python types: they permit fewer behaviors statically to enable the unboxing. Operations such as `and` (boolean conjunction), for example, do not accept a mix of primitive and non-primitive arguments. These primitive types are another central aspect of gradual soundness.

Because performance is the top priority for primitive types, they are intentionally incompatible with the dynamic type. Casts between a primitive type and the dynamic type are rejected statically — another violation of the gradual guarantee — to encourage the spread of primitives in typed code. At the boundaries to untyped Python code, however, Static Python inserts casts. These casts let programmers introduce primitive types in one module without having to rewrite its untyped dependents.

### 2.6 Behavioral Changes to Python

Relative to Python, Static Python makes a few behavioral changes to enable sound types and strong optimizations: module-level variables cannot be mutated; class attributes must be set in an __init__ method to ensure predictable layouts (all PEP 484 checkers agree on this); typed attributes are resolved statically, and thus cannot be overridden with descriptors or properties; and multiple inheritance of either typed classes or untyped children of typed classes is forbidden.

Several other Python features are unsupported at present: enums; the @dataclass decorator; overrides of the __setattr__ and __getattr__ methods; keyword arguments with default expressions (rather than values); and calls to keyword functions that use **kwargs with either a custom object or a dict with extra keys. The Static Python team plans to lift these restrictions in the future.

### 2.7 Gradual Soundness

Static Python represents a new approach to the design of gradually-typed languages. Whereas prior works choose to either force changes upon untyped code with *concrete* types [39, 67] or accommodate untyped code as-is with *migratory* types [52, 62], Static Python provides a multi-level type system that supports both styles without expensive run-time checks. Shallow types accommodate Python code, but offer little in the way of type-based reasoning and type-directed optimization. Concrete types and primitive types give stronger type guarantees, but have drawbacks: concrete types force changes upon untyped code and primitive types add restrictions to typed code. The overall system, in which programmers can choose between untyped code and several increasingly-restrictive typed languages, is a combination of gradual typing and progressive types [44] that we call *gradual soundness*.

Compared to traditional gradual typing, gradual soundness is a refinement. The original gradual typing vision is to find a best-of-both-worlds combination of typed and untyped code [24, 35, 52, 60]. Gradual soundness fits under this broad umbrella, and contributes the idea that multiple type systems may be needed for an optimal mix. Gradual typing in the refined sense [54] calls for a dynamic type that satisfies the gradual guarantees. A gradually-sound language may choose to satisfy these guarantees. Static Python, however, does not (Section 2.2).

## 3 Runtime System Highlights

The Static Python runtime system, formally known as Cinder, extends CPython 3.8 in several ways to take full advantage of static types. Cinder includes tailored bytecode instructions, virtual method tables, concrete datatypes, a registry of typed modules, and a JIT compiler. These main ingredients of Cinder may be of interest to other teams seeking to add sound gradual typing to an existing language. For example, Node developers may wish to fork V8 and experiment with bytecode instructions that leverage sound static types.

**Bytecode Instructions** All CPython 3.8 instructions have the same behavior in Cinder. Cinder adds instructions to help with one of three tasks: expressing runtime checks, initializing concrete-typed values, or efficiently performing a standard action. As an example of the third kind, the `FAST_LEN` instruction quickly computes the length of a built-in value, which helps to optimize loops. Static Python uses type information to decide where this instruction is appropriate and applies it as an optimization.

**Virtual Method Tables** Cinder adds virtual method tables (vtables) to classes. These tables help to speed up method dispatch relative to Python's dynamic lookup. Calls to static methods that appear in statically-typed code use the vtable to find an address for the method. If the method is part of a final class, then the call is further optimized to a direct function-call jump. (Both vtable lookup and direct jumps are supported by Cinder-specific bytecode instructions.)

The implementation of vtables is built on the Python vectorcall protocol; it is not a from-scratch development. Cinder vtables are further specialized to check untyped overrides of typed methods using a wrapper implemented in C (rather than in Python) to reduce performance costs.

**Concrete Data Structures** The concrete versions of built-in data structures come with both a type and an implementation. The implementation provides the same interface as the built-in, but uses type tags to reject certain operations.

For example, the type CheckedDict$[K, V]$ describes a concrete dictionary with keys of type $K$ and values of type $V$. The implementation has three main components:

- The constructor requires two type parameters and a Python dictionary (a Dict). It checks that the elements of the dictionary match the key and value types.
- All untyped writes must be protected by casts. Every operation that mutates or extends a checked dictionary must validate any untyped arguments that it receives.
- The dictionary value stores a type tag to support casts from the dynamic type. A tag represents a three-part type CheckedDict$[K, V]$ and is implemented as a pointer into a global registry of instantiated types. When a concrete dictionary enters typed code from an untyped context, its tag must match the expected type exactly. For example, type CheckedDict$[K, V]$ matches the type CheckedDict$[K', V']$ only if $K$ is equal to $K'$ and $V$ is equal to $V'$.

In general, other checked datatypes have the same three components: a constructor, a set of checked update functions, and a tag. The tag is supported by a global registry of instantiated concrete types.

**Classloader** Static Python keeps track of typed functions and typed classes with a specialized classloader. At runtime, the classloader keeps a registry of typed objects. The registry helps the bytecode reference objects and types via their module names.

**Method-at-a-Time JIT** Cinder includes a JIT (just-in-time) compiler for its bytecode. Programmers can enable the JIT by supplying a list of method names to the compiler.

Any method can be JIT-compiled whether or not it uses Static Python types, though fewer types usually result in fewer optimizations.

## 4 Model

To validate the design of Static Python, we developed a model of the language in PLT Redex [28]. The model covers a substantial part of the Python language, including assertions, loops, exception handlers, and delete statements. It follows Static Python's approach to typing these features. The model is on GitHub (accessed 2022-05-17):

https://github.com/brownplt/insta-model

Because there is no prior formalization of (all of) Python and of Static Python, we cannot *verify* that the model matches them. Instead, we have applied three methods to give confidence that our model matches reality. First, we manually reviewed the model in depth—using our own expertise—to look for non-conformance. Second, we used Redex's property-based testing tools [29] to check type soundness. Specifically, we used Redex to generate well-formed terms (using its default algorithm) and checked whether the well-typed terms that terminated within a fixed time budget (1,200 expressions and 11,000 programs in a typical run) produced only well-typed values and allowed errors. Finally, we employed the well-established method of testing end-to-end conformance with a test suite [5, 14, 26, 42, 43]. We translated 265 tests from the Static Python regression suite to the syntax of the model and confirmed that the results do match, which suggests that the model conforms to actual Static Python.

For most of the 265 tests, the translation is automatic. A few tests required hand-pruning to remove features that the model does not handle (52 total). Static Python has 537 other tests (802 total) that we did not use because they fall outside the scope of the model. Section 5 summarizes the out-of-scope features and Appendix C gives a detailed categorization.

**The Payoff: Issues Reported** The modelling process helped uncover several critical issues in Static Python. This is a very important payoff given that Static Python is running in production at Instagram. Overall, we made 26 bug reports (Appendix A) to the Static Python team via GitHub. Five of these were soundness issues, one of which we could exploit to raise a segmentation fault. All of these soundness bugs have been fixed. Of the remaining issues, five were relatively minor; these dealt with confusing error messages and incorrect tests. The remaining 16 issues report bugs in language design and implementation. The Static Python team has acknowledged these as bugs by applying a specific GitHub label: `sp-correctness`.

To give one example issue, Static Python incorrectly accepted the following program whereas our model reported a type error:

```
from typing import Optional

def f(x: Optional[str]) -> str:
  while True:
    if x is None:
      break
    return x
```

The function `f` expects either a string or the `None` value and promises to return a string. When called with `None`, however, the function breaks out of the `while` loop and implicitly returns `None` contrary to the return type. Static Python had failed to account for the `break` and implicit return. More concerningly, the associated test case *expected* the program to type check. Our model caught this specification error.

**Section Outline** The rest of this section illustrates our Redex model using a tiny formalization to highlight the boundaries between typed and dynamic-typed code. Section 4.1 presents a surface syntax. Section 4.2 explains how types get enforced at run-time. Section 4.3 argues that the overall approach toward boundaries is sound.

### 4.1 Surface Syntax and Types

Figure 3 presents a core syntax for programs. A program is a sequence of statements; a statement defines a variable, function, or class. These definitions may only appear on the top level and they all require type annotations. Functions must have one positional argument. Classes must declare one parent (either Object or another class), one field, and one method. Expressions describe values and simple computations. The basic values are the none value, integers, booleans (which are the integers 0 and 1), and strings. There are two data structures: dictionaries and checked dictionaries (Section 2.3). Other expressions describe function calls, dictionary reads and writes, object creation, object field reads and writes, and method calls.

In Python, the syntax for expressions overlaps with the syntax of types. For example, the Python name `None` is both an expression and a type, and the Static Python names `chkdict` and `CheckedDict` are synonyms. Figure 3 does not follow these standards. Instead, it keeps the two syntaxes distinct. Expressions use lowercase names (`none`, `chkdict`) and types use capitalized names (`None`, `CheckedDict`).

Surface types S include the dynamic type, types for the basic values, one type C for every user-defined class, and union types. We choose to write the dynamic type as Dyn for simplicity; technically, Static Python does not have a surface syntax for its dynamic type and instead allows missing types (Section 2.2). We assume that all unions are written in a flat and simplified form, e.g., that $\mathsf{Union}[\mathsf{Int}, \mathsf{Union}[\mathsf{Dyn}, \mathsf{C}_0]]$ would be normalized to Dyn.

Static Python does not enforce all surface types against untyped code, only the evaluation types presented in the next subsection (Section 4.2). For this reason, we omit the surface typing judgment, which is a kind of linter that merely scans typed code for logical errors.

$$\begin{array}{lcl}
\textit{prog} & = & \textit{stmt} \mid \textit{stmt},\ \textit{prog} \\[4pt]
\textit{stmt} & = & \mathsf{x{:}S} = \textit{expr} \mid \mathsf{def\ f(x{:}S)\ {\text{->}}\ S} : \textit{expr} \mid \\
 & & \mathsf{class\ C(C):\ x{:}S} = \textit{expr};\ \mathsf{def\ f(self,x{:}S)\ {\text{->}}\ S} : \textit{expr} \\[4pt]
\textit{expr} & = & \mathsf{x} \mid \mathsf{none} \mid \textit{int} \mid \textit{bool} \mid \textit{str} \mid \mathsf{f}(\textit{expr}) \mid \\
 & & \{\mathsf{x} : \textit{expr},\ldots\} \mid \mathsf{chkdict[S,S]}(\{\mathsf{x} : \textit{expr},\ldots\}) \mid \\
 & & \textit{expr}[\textit{expr}] \mid \textit{expr}[\textit{expr}] = \textit{expr} \mid \\
 & & \mathsf{C}(\textit{expr},\ldots) \mid \textit{expr}.\mathsf{x} \mid \textit{expr}.\mathsf{x} = \textit{expr} \mid \textit{expr}.\mathsf{f}(\textit{expr}) \\[4pt]
\mathsf{S} & = & \mathsf{Dyn} \mid \mathsf{None} \mid \mathsf{Int} \mid \mathsf{Bool} \mid \mathsf{Str} \mid \mathsf{C} \mid \\
 & & \mathsf{Dict[S,S]} \mid \mathsf{CheckedDict[S,S]} \mid \mathsf{Union[S,\ldots]} \\[4pt]
\Gamma & = & \cdot \mid \mathsf{x{:}S},\ \Gamma \mid \mathsf{f(S)\ {\text{->}}\ S},\ \Gamma \mid \mathsf{class\ C(C):\ x{:}S;\ f(C,S)\ {\text{->}}\ S},\ \Gamma \\[4pt]
\mathsf{x,f} & = & \text{variable names}
\end{array}$$

Abbreviation: $\mathsf{Optional[S]} = \mathsf{Union[None,S]}$

Convention: the set of all class types $\mathsf{C}$ contains the top type $\mathsf{Object}$

**Figure 3** Surface syntax and types

$$\begin{array}{lcl}
\mathsf{T} & = & \mathsf{Dyn} \mid \mathsf{None} \mid \mathsf{Int} \mid \mathsf{Bool} \mid \mathsf{Str} \mid \mathsf{C} \mid \\
 & & \mathsf{Dict} \mid \mathsf{CheckedDict[T,T]} \mid \mathsf{Optional[T]}
\end{array}$$

$$R(\mathsf{S_0}) = \begin{cases}
\mathsf{Dict} & \text{if } \mathsf{S_0} = \mathsf{Dict[S,S]} \\
\mathsf{Optional}[R(\mathsf{S_1})] & \text{if } \mathsf{S_0} = \mathsf{Optional[S_1]} \\
\mathsf{Dyn} & \text{if } \mathsf{S_0} = \mathsf{Union[S,\ldots]} \text{ and } \mathsf{S_0} \neq \mathsf{Optional[S]} \\
\mathsf{CheckedDict}[R(\mathsf{S_1}),R(\mathsf{S_2})] & \text{if } \mathsf{S_0} = \mathsf{CheckedDict[S_1,S_2]} \\
\mathsf{S_0} & \text{otherwise}
\end{cases}$$

**Figure 4** Evaluation types, Surface-to-Evaluation mapping

Relative to Static Python and our Redex mechanization, the formalization in Figure 3 omits many details of Python including class variables, imports, conditionals, and exception handlers. These details are crucial in the mechanization, which tests whether Static Python soundly approximates Python. They are less important here, where our focus is on type boundaries. None of the omitted features give substantially new ways for typed code to interact with dynamic code.

### 4.2 Evaluation Types, Casts, and Typing Judgment

Evaluation types $\mathsf{T}$ are a simplification of the surface types. These are the types that Static Python promises to soundly enforce. Figure 4 presents the syntax of evaluation types and a retraction $R(\cdot)$ from surface types to evaluation types. As the retraction shows, the evaluation types make two main changes:

1. The parameterized type for Python dictionaries $\mathsf{Dict[S_0,S_1]}$ gets replaced with a raw type $\mathsf{Dict}$. The raw type behaves the same as $\mathsf{Dict[Dyn,Dyn]}$ would.

**Table 2** Description of run-time checks for the evaluation types

| Eval. Type T | Run-time check |
|---|---|
| $\mathsf{Optional}[\mathsf{T}_0]$ | Accepts either the none value or values that match $\mathsf{T}_0$ |
| Dict | Accepts any Python dictionary |
| $\mathsf{CheckedDict}[\mathsf{T}_0, \mathsf{T}_1]$ | Accepts checked dictionaries parameterized by $\mathsf{T}_0$ and $\mathsf{T}_1$ |
| Dyn | No check needed, accepts any value |
| $\mathsf{C}_0$ | Accepts instances of the class $\mathsf{C}_0$* |
| None | Accepts the Python none value |
| Int | Accepts integers and booleans |
| Bool | Accepts booleans |
| Str | Accepts strings |

* In full Static Python, the check for Object accepts all values except for primitives (Section 2.5)

2. Unions get replaced with the dynamic type, except for the special case of unions with none ($\mathsf{Optional}[\mathsf{S}_0]$).

#### 4.2.1 Casts

Static Python evaluation types can all be enforced fully and efficiently with casts. Table 2 describes the casts in detail. Most call for a tag check, i.e., a Python isinstance test. The sole exception is optional types, which require a tag check and a test for the none value. Checked dictionaries also rely on tag checks; a checked dict type with keys $\mathsf{T}_0$ and values $\mathsf{T}_1$ accepts only checked dict values that were initialized with exactly the same key and value type. As noted in Section 2.2, this exact-match rule applies even when $\mathsf{T}_0$ or $\mathsf{T}_1$ is the dynamic type.

No cast requires traversing a data structure. Similarly, no cast allocates a wrapper to check higher-order behaviors in a delayed fashion. This latter property means that blame-tracking [12] is trivial because casts either succeed or fail immediately. In other words, Static Python provides the same *immediate accountability* as Nom [39].

#### 4.2.2 Expression Typing, Cast Insertion

To show where Static Python needs to insert casts, we present a selection of typing rules. Recall that a program declares variables, functions, and classes (Figure 3). These declarations fill a type environment ($\Gamma$). Relative to the current environment, the expressions within each statement must satisfy the typing judgment in Figure 5.

By contrast to our Redex model, the typing judgment outlined in Figure 5 *does not describe* a typechecking algorithm. It is rather a declarative set of rules chosen to illustrate two points: how Static Python reasons about data structure types and how it enforces promises made via the dynamic type. The key steps that the Redex model takes to turn these rules into an algorithm is to remove the two coercion rules (both are marked with asterisks: C-SUB* and MATR*) and to allow restricted coercions at elimination forms. For example, a function application may apply either consistent subtyping or materialization to its argument. It cannot apply both kinds of coecion because that can be abused to delay any static error until run-time.

$\boxed{\Gamma \vdash_{\mathrm{E}} expr : \mathsf{T}}$ selected rules

$$\textsc{S-App}\ \frac{\mathsf{f}_0\,(\mathsf{T}_0)\ \text{->}\ \mathsf{T}_1 \in \Gamma \qquad \Gamma \vdash_{\mathrm{E}} expr_0 : \mathsf{T}_0}{\Gamma \vdash_{\mathrm{E}} \mathsf{f}_0\,(expr_0) : \mathsf{T}_1} \qquad \textsc{D-App}\ \frac{\mathsf{f}_0 : \mathsf{Dyn} \in \Gamma \qquad \Gamma \vdash_{\mathrm{E}} expr_0 : \mathsf{Dyn}}{\Gamma \vdash_{\mathrm{E}} \mathsf{f}_0\,(expr_0) : \mathsf{Dyn}}$$

$$\textsc{D-Set}\ \frac{\Gamma \vdash_{\mathrm{E}} expr_0 : \mathsf{Dict} \qquad \Gamma \vdash_{\mathrm{E}} expr_1 : \mathsf{Dyn} \qquad \Gamma \vdash_{\mathrm{E}} expr_2 : \mathsf{Dyn}}{\Gamma \vdash_{\mathrm{E}} expr_0[expr_1] = expr_2 : \mathsf{None}}$$

$$\textsc{CD-Set}\ \frac{\Gamma \vdash_{\mathrm{E}} expr_0 : \mathsf{CheckedDict}[\mathsf{T}_0, \mathsf{T}_1] \qquad \Gamma \vdash_{\mathrm{E}} expr_1 : \mathsf{T}_0 \qquad \Gamma \vdash_{\mathrm{E}} expr_2 : \mathsf{T}_1}{\Gamma \vdash_{\mathrm{E}} expr_0[expr_1] = expr_2 : \mathsf{None}}$$

$$\textsc{F-Set}\ \frac{\Gamma \vdash_{\mathrm{E}} expr_0 : \mathsf{C}_0 \qquad \mathsf{x}_0 : \mathsf{T}_0 \in \Gamma^*\,(\mathsf{C}_0) \qquad \Gamma \vdash_{\mathrm{E}} expr_1 : \mathsf{T}_0}{\Gamma \vdash_{\mathrm{E}} expr_0.\mathsf{x}_0 = expr_1 : \mathsf{None}}$$

$$\textsc{M-App}\ \frac{\Gamma \vdash_{\mathrm{E}} expr_0 : \mathsf{C}_0 \qquad \mathsf{f}_0\,(\mathsf{C}_0, \mathsf{T}_0)\ \text{->}\ \mathsf{T}_1 \in \Gamma^*\,(\mathsf{C}_0) \qquad \Gamma \vdash_{\mathrm{E}} expr_1 : \mathsf{T}_0}{\Gamma \vdash_{\mathrm{E}} expr_0.\mathsf{f}_0\,(expr_1) : \mathsf{T}_1}$$

$$\textsc{C-Sub}^*\ \frac{\Gamma \vdash_{\mathrm{E}} expr_0 : \mathsf{T}_1 \qquad \Gamma \vdash \mathsf{T}_1 \sqsubseteq \mathsf{T}_0}{\Gamma \vdash_{\mathrm{E}} expr_0 : \mathsf{T}_0} \qquad \textsc{Matr}^*\ \frac{\Gamma \vdash_{\mathrm{E}} expr_0 : \mathsf{T}_1 \qquad \mathsf{T}_1 \prec \mathsf{T}_0}{\Gamma \vdash_{\mathrm{E}} expr_0 : \mathsf{T}_0}$$

Where $\Gamma^*\,(\mathsf{C})$ collects field and method types from C and from its parents

$\boxed{\Gamma \vdash \mathsf{T} \sqsubseteq \mathsf{T}}$

$$\frac{\Gamma \vdash \mathsf{T}_0 <: \mathsf{T}_1}{\Gamma \vdash \mathsf{T}_0 \sqsubseteq \mathsf{T}_1} \qquad \frac{}{\Gamma \vdash \mathsf{T}_0 \sqsubseteq \mathsf{Dyn}}$$

$\boxed{\mathsf{T} \prec \mathsf{T}}$

$$\frac{}{\mathsf{Dyn} \prec \mathsf{T}_0}\ \mathsf{T}_0 \neq \mathsf{Dyn}$$

$\boxed{\Gamma \vdash \mathsf{T} \leq: \mathsf{T}}$ reflexive, transitive closure of the following <: relation:

$$\frac{}{\Gamma \vdash \mathsf{T}_0 <: \mathsf{Object}} \qquad \frac{}{\Gamma \vdash \mathsf{Bool} <: \mathsf{Int}} \qquad \frac{\Gamma \vdash \mathsf{T}_0 <: \mathsf{T}_1}{\Gamma \vdash \mathsf{Optional}[\mathsf{T}_0] <: \mathsf{Optional}[\mathsf{T}_1]}$$

$$\frac{}{\Gamma \vdash \mathsf{None} <: \mathsf{Optional}[\mathsf{T}_0]} \qquad \frac{\Gamma \vdash \mathsf{T}_0 <: \mathsf{T}_1}{\Gamma \vdash \mathsf{T}_0 <: \mathsf{Optional}[\mathsf{T}_1]} \qquad \frac{\text{class } \mathsf{C}_0\,(\mathsf{C}_1) :\ \ldots;\ \ldots \in \Gamma}{\Gamma \vdash \mathsf{C}_0 <: \mathsf{C}_1}$$

**Figure 5** Expression typing, consistent subtyping, and materialization

The first two typing rules are for function application. A typed function may be applied to an argument that matches its domain type, in which case it computes a value that matches its codomain type. A dynamically-typed variable may be applied to any input to yield a dynamically-typed result. The following rules illustrate writes to Python dictionaries and checked dictionaries. A shallow Python dict may be updated with any kind of key and value. By contrast, a concrete checked dict requires keys and values that match its type parameters. Next we have two rules for classes: F-SET says that writes to a class field must match the declared field type; and M-APP shows that typed methods impose similar constraints as typed functions.

The final two rules, C-SUB* and MATR*, depend on auxiliary judgments for consistent subtyping ($\sqsubseteq$) and materialization ($\prec$). Consistent subtyping relates type $T_0$ to type $T_1$ if they are related by static subtyping ($\leq:$) or if $T_1$ is the dynamic type. Materialization relates the dynamic type to any non-dynamic type. Occurrences of the MATR* rule are downcasts that require a run-time check.

The name *materialization* and the rule MATR* are inspired by prior work [6]. By itself, the judgment is merely an upside-down type precision relation [15, 53]. Combined with the typing rule, however, materialization is a concise way to find where a well-typed program needs casts to ensure soundness. For example, suppose that f is a function from integers to the dynamic type and that x is a variable with the dynamic type. An application f(x) can satisfy the type Str using two materializations:

$$\textsc{Matr*}\ \dfrac{\textsc{Matr*}\ \dfrac{\dfrac{\Gamma \vdash_{\mathbf{E}} \mathsf{x} : \mathsf{Dyn} \qquad \mathsf{Dyn} \prec \mathsf{Int}}{\Gamma \vdash_{\mathbf{E}} \mathsf{x} : \mathsf{Int}}}{\Gamma \vdash_{\mathbf{E}} \mathsf{f(x)} : \mathsf{Dyn}} \qquad \mathsf{Dyn} \prec \mathsf{Str}}{\Gamma \vdash_{\mathbf{E}} \mathsf{f(x)} : \mathsf{Str}}$$

$$\text{Where } \Gamma = \mathsf{x : Dyn, f(Int)\text{ -> }Dyn}$$

Consequently, this derivation calls for two casts at the MATR* applications.

We end here, with materialization, rather than present a semantics for the formalization. After all, the main benefit of a full formal semantics is to validate the behavior of complex expressions—and this job is better left to the Redex mechanization, which supports a much larger subset of Python (Section 4.1).

Furthermore, our casts-via-materialization rule assumes that the type checker has access to the whole program. In practice, Static Python approximates our ideal casts in two steps to support interactions with unanalyzed Python modules. First, Static Python inserts casts in an overapproximate way:

- typed functions and methods begin by checking all their arguments;
- typed classes check all field writes;
- concrete-typed dictionaries check all key and value writes; and
- typed code checks the results of function calls, references, and other elimination forms whenever there is a materialization from the dynamic type.

Second, the type-directed optimizer skips these casts wherever a well-typed context interacts with a typed value. Section 5 explains the optimizer in more detail.

### 4.3 Type Boundary Soundness

The soundness of our model is easily seen by reduction to Nom, a gradual language that allows fine-grained interactions with untyped code and comes with a detailed proof of soundness [39, 40]. Given a program written in the syntax of Figure 3, one can derive a Nom program with the same type boundaries by replacing every checked dict instance with a fresh class type. Because distinct checked dict types are incompatible, the original boundaries impose the same constraints as the translated Nom ones.

For interested readers, we offer the following direct argument as well. Boundaries to less-typed code would be *unsound* if an untyped (or partially-untyped) value could enter a typed context without a validating check. The question is thus whether all boundaries are properly guarded. Our answer has two parts. First, observe that every evaluation type comes with a decidable cast (Table 2). For basic types such as strings, a tag check clearly suffices. For parameterized types such as $\mathsf{CheckedDict}[\mathsf{T}_0, \mathsf{T}_1]$, the exact-match semantics is also decidable (and also implemented with tag checks). Second, observe that boundaries can only relate the dynamic type to an evaluation type; there are no significant boundaries that relate a partially static type to an evaluation type. This is due to the exact-match semantics for checked types and the fact that Optional[Dyn] normalizes to the dynamic type. Therefore, inserting casts at static occurrences of the materialization rule is enough to protect all boundaries.

## 5 Scaling to Python

The Redex model intentionally does not cover all of Python. Some aspects of Static Python are left out because they are straightforward to handle. These include primitive types (Section 2.5), module-level variables, and the boundary to untyped Python (Section 5.1). Other Python features are left out because their Static Python semantics is identical to standard Python (Section 5.1). Our model also does not cover the optimizer (Section 5.3) because a formal account of type-directed optimization would require substantial additional components; namely, a model of the Static Python bytecode and a faithful rendering of its transformations.

### 5.1 Interactions with Open-World Python Code

Although Static Python is technically a new language, it is designed for gradual adoption. Python programmers should be able to add types module-by-module to an existing codebase. Consequently, Static Python supports interactions with Python modules by letting the Python code run with zero static constraints and minimal dynamic constraints.

There are two cases in which Static Python types impose dynamic constraints on untyped Python code. First, concrete types (e.g., CheckedDict) reject inputs that were not created via a checked constructor. If a client of a Python module decides to impose concrete types, the Python module may need to change. Second, typed modules prevent updates to module-level variables.

Interactions with open-world [66] Python code pose a minor threat to soundness because typed functions (and methods) cannot assume that their arguments are well-typed. All arguments sent from typed contexts get validated either statically or via materialization casts, but arguments from untyped contexts are unchecked. For this reason, Static Python compiles every typed function to check its inputs. The optimizer can bypass these checks for typed-to-typed calls (Section 5.3). Similarly, concrete types such as CheckedDict must check writes that come from untyped contexts.

### 5.2 Dynamic Python Features

Static Python does not ascribe types to the following Python features. These are not covered in the model because the implementation simply assigns the dynamic type and lets the runtime treat them as untyped Python code. For each, the dynamic type is a reasonable choice; accurate static types would be burdensome to maintain.

**First-Class Classes** Static Python does not attempt to type first-class classes: partly because they have yet to appear in performance-critical code in the Instagram web server and partly because they do not fit well with a nominal type system. The straightforward but restrictive approach for first-class nominal class types is to force code that uses a first-class class to expect subtypes of a particular named static class. Flatt et al. [13] propose a more flexible approach, but it requires a second layer of *interface types* atop the nominal hierarchy. MonNom [40] uses interfaces in a similar way to accommodate structural objects.

On a related note, Static Python has no support for first-class functions or for structural types as defined by PEP 544 [31]. Interfaces may be a promising way to support these types.

**Multiple Inheritance** Python allows classes to inherit from a list of parents. Static Python accommodates this behavior provided that at most one parent is typed (or uses __slots__ to specify its layout; plain Python has the same restriction).

Static Python does not leverage types for classes with multiple parents because of vtable layout issues. Efficient vtable dispatch requires that each method name resides in a fixed slot across all subclasses. With multiple inheritance, however, parents may disagree about which method lives in each slot. Similarly, Static Python cannot optimize instance variable reads because their slots may have a conflict among parents.

**Dynamic Execution** Calls to eval and exec have the dynamic type. Studies of JavaScript and R have shown that many uses of dynamic execution can be removed through simple adjustments [17, 36, 48]. Assuming these findings carry over to Python, similar adjustments would be preferable over attempting to type eval.

**Table 3** Representative Cinder bytecode instructions

| Instruction | Purpose | Description |
|---|---|---|
| CAST | Soundness | Assert that a value matches a type |
| CHECK_ARGS | Soundness | Cast all inputs to a function |
| BUILD_CHECKED_MAP | Constructor | Make a checked dictionary |
| TP_ALLOC | Constructor | Make a Static Python object |
| INVOKE_FUNCTION | Optimization | Execute a direct function call |
| INVOKE_METHOD | Optimization | Execute a method via the object's vtable |
| LOAD_FIELD | Optimization | Read an object field |
| STORE_FIELD | Optimization | Write to an object field |
| FAST_LEN | Optimization | Get the length of a built-in value |
| REFINE_TYPE | Optimization | Type declaration for the JIT |

### 5.3 Bytecode and Optimizations

Static Python can generate efficient code because it targets the Cinder runtime. Cinder provides bytecode instructions that check types at run-time, construct Static Python data structures, and perform type-directed actions. Table 3 lists a few representative instructions. The two instructions that express run-time checks are CAST and CHECK_ARGS. The former checks a value against a type. The latter is for functions and methods; it checks all inputs to a function against their declared types. The instructions BUILD_CHECKED_MAP and TP_ALLOC allocate Static Python-specific data structures. The first creates a checked dictionary; the second allocates an uninitialized instance of an object (Section 3). Cinder comes with a similar instruction to build checked lists. The remaining instructions are for optimization. Both INVOKE_FUNCTION and INVOKE_METHOD are alternatives to Python's dynamic call dispatch. The former uses the classloader to quickly find the address of a function; the latter uses a vtable lookup to find a method. The load and store instructions for fields improve upon Python's generic attribute lookup. In the JIT, these instructions can be further optimized to a single assembly instruction. Lastly, the REFINE_TYPE instruction tells the JIT about the type of a local value when it is not clear from the context.

**Reducing Casts** In addition to upgrading bytecode instructions to optimized ones, Static Python takes care to minimize the type casts that it executes at run-time. In other words, it takes care to slow code down as little as possible.

Part of this goal is met by inserting casts only in positions where the dynamic type flows into a static type. The materialize rule in the model illustrates this policy, which ensures that well-typed writes to classes and to concrete dictionaries have no cost. Typed functions, however, are compiled with a CHECK_ARGS instruction that defensively casts all arguments. By convention, this instruction always appears on the first line of a typed function body. The optimizer uses this convention to skip argument checks by jumping past them when it is safe to do so; that is, whenever a typed function or method gets called in a typed context.

## 6 Production Experience

The Instagram web server has been using Static Python code to serve requests in production since April 2021. Overall, the results are very encouraging. Instagram's internal profiling tools attribute a 3.7 % improvement in CPU efficiency (Section 6.1) to Static Python conversions (Section 6.2). This is a big improvement at Instagram scale.

Developers have converted over 500 modules to Static Python thus far. Despite some initial regressions, none of the converted modules ran slower after small rewrites (Section 6.3) and/or enabling the JIT on certain functions. Only 9 modules use concrete types and only 12 modules use primitive types, but these features delivered critical performance gains in key modules.

### 6.1 How to Interpret the CPU Efficiency Result

The Instagram web server code runs on a large number of host machines, each of which continuously handles requests from a common pool. Improvements to the server codebase should make these machines more efficient as they handle arbitrary requests.

To measure the effect of a code change on these hosts, an internal profiling tool selects two representative sets of machines to run as experimental and control groups. The experimental group gets the latest version of the server code; the control group gets the previous version. Next, the profiling tool slowly increases the number of requests that these machines receive until each is running at maximum load. Once the machines are fully allocated, the tool measures how many requests per second each group is able to handle. If the experimental group can serve more requests per second, then the change is a success. Assuming that the experimental and control groups have equivalent hardware capabilities, and assuming that both groups receive a representative sample of requests, then the results of an experiment (increase/decrease in CPU efficiency) should predict actual performance.

The one caveat with this CPU efficiency measure is that it is pinned to a specific point in time. Changes in product code and in client behavior can change the size of a typical web server request, which changes the number of requests that a server can handle per second. Taking the sum of several CPU efficiency changes over a long timespan (as we have done) is therefore only a rough measure of their net effect.

### 6.2 Migration Path

Static Python first entered the web server codebase as a replacement for a few critical Cython modules. Cython had improved performance by compiling these modules to C, but its partial adoption led to an awkward workflow with a few compiled modules spread across an interpreted codebase. Replacing these modules with Static Python let programmers return to a conventional Python workflow. Furthermore, Static Python types combined with Cinder resulted in a 0.7 % improvement in web server CPU efficiency. Primitive types (Section 2.5) were essential for matching Cython.

Later Static Python migrations have been directed by profiling to find frequently-executed code. During the first half of 2021, the Static Python team identified hot modules and proposed types plus small code changes to the maintainers of these modules (Section 6.3). The accepted proposals resulted in a 1 % improvement in CPU efficiency. During the second half of 2021, the Static Python team applied the same process at a larger scale. They also modified a code-generating module to output Static Python code; this one change resulted in over four hundred generated typed modules. Overall, the second-half changes added 2 % to the number of requests that servers could handle per second at maximum CPU load.

As of December 2021, the Static Python team has converted 541 modules. Most of these came from the code-generating tool (417); the rest are from hand conversions (124). These modules frequently interact with untyped modules in the codebase. According to an analysis of identifiers that cross module-dependence boundaries, over 30,000 identifiers go between typed and untyped code. Two-thirds of these crossings are exports from Static Python modules to untyped modules; in other words, typed identifiers are widely used throughout the web server.

### 6.3 Analyzing Code Changes

Because the Static Python team has been changing code as well as types during its migrations, the question arises as to whether the refactoring caused more speedups than the types and gradual soundness. We therefore reviewed 30 conversion diffs that significantly improved performance to assess the extent of the code changes. Across the diffs, there are nine common kinds of changes: five that satisfy the type checker and four that aim to improve performance.

#### 6.3.1 Code Changes for the Type Checker

1. *Fix type errors due to mock wrappers*. When test code uses a mock wrapper, it may change the return type of a function because wrappers return a `MagicMock` object by default. The fix is to specify a return value.
2. *Change mocked function to expect positional arguments*. Static Python currently rewrites the bytecode for all typed-to-typed function calls to use positional arguments. Mock-wrapped functions therefore need to use positional arguments.
3. *Organize class and instance attributes*. Whereas Python allows class attributes to serve as default values for instance attributes, Static Python does not.
4. *Move unsupported Python features*. Code that uses the unsupported features listed in Section 2.6 must be changed or moved to an untyped module.
5. *Refine some Pyre annotations*. Well-typed Pyre code is not always well-typed Static Python code. Appendix B lists the main points of friction.

**Table 4** Microbenchmark performance ratios

| Name | Typed / Python | Refined / Python |
|---|---|---|
| deltablue | 0.59 | 0.30 |
| fannkuch | 1.03 | 0.46 |
| nbody | 1.09 | 0.24 |
| richards | 0.53 | 0.22 |

#### 6.3.2 Code Changes for Performance

1. *Use primitive types*. Integers and booleans in hot code paths run fastest with primitive types (Section 2.5). Twelve modules currently use primitives.
2. *Use concrete types*. Three modules currently use a concrete dictionary type (of the form CheckedDict[$K, V$]) and six modules use a checked list type.
3. *Change functions to accept positional arguments*. Functions calls that use only keyword arguments are not yet optimized by Static Python.
4. *Convert @classmethod to @staticmethod*. Static methods can get invoked directly as functions, bypassing the class vtable. Twenty-four modules use static methods (many of these pre-date Static Python).

#### 6.3.3 Conclusion

In general, the patches responsible for performance gains contain minor code changes and affect test code rather than production code. We conclude that the addition of Static Python types to a codebase can significantly improve performance without major refactoring, and that refactoring is not the main reason for Instagram's speedups.

### 6.4 Microbenchmarks

Table 4 compares Static Python to Python on public microbenchmarks [59]. These microbenchmarks are admittedly small and quite different than typical application code, but they provide an auditable and reproducible way to measure performance.

Each row presents two ratios. The first compares a simply-typed version of the benchmark to untyped Python code. The second compares a hand-refined benchmark to untyped Python. The refinements are modest, mostly adding primitive types to hot loops. All runtimes for this table came from a 64-bit CentOS Linux system using a build of Static Python and Cinder compiled with default settings. Each benchmark version ran 11 times total: once to warm up the bytecode caches (.pyc files) and 10 times to compute an average runtime (using the time utility). Every benchmark version invoked the JIT on the same set of functions. Appendix D presents further details.

Almost all ratios are less than one, showing that Static Python achieves a speedup over Python (typed avg. 0.8, refined avg. 0.3). The exceptions are the unrefined versions of fannkuch and nbody, which is because these benchmarks focus on number-crunching and Static Python optimizes arithmetic only for primitive types. Changing a few types in fannkuch to primitives gives an immediate speedup (ratio = 0.90).

**Threat to Validity** Table 4 does not analyze benchmark configurations that mix typed and untyped code. A study of mixed configurations (e.g., [22, 23]) may reveal performance bugs that have not appeared in production.

## 7 Related Work

Concrete types originated in Thorn [4, 67] and StrongScript [49], two early languages that support typed and untyped objects. Thorn is a scripting language for the JVM; StrongScript is designed to interoperate with JavaScript. Concrete types came to Static Python by way of Nom [39] and its successor MonNom [40]. The impressive performance figures in the original Nom paper inspired the Static Python team to adapt the method. But whereas MonNom has been exploring designs for generic concrete types that support (a flexible form of) the gradual guarantees [54], Static Python opts for a simpler and more rigid design (Section 2.2).

With its mix of shallow and concrete types, Static Python is one of a few gradual language that lets programmers choose between multiple type-enforcement strategies. Thorn and StrongScript offer a choice of concrete types and optional (unsound) *like* types. Typed Racket supports deep (structural) types, shallow types, and optional types [20]. By contrast to these others, Static Python asks programmers to change their code in way that reflects changes to the types (Section 2.3). The changes impose a maintenance burden, but enable performance improvements and an efficient implementation of run-time checks.

Many aspects of our Redex model for Static Python are adapted from the Full Monty core calculus for Python [43]. In particular, the Full Monty paper reminded us that booleans are integers in Python; Static Python had overlooked this detail.

Production experience with Static Python suggests that a method-based JIT can eliminate the costs of shallow types. This finding complements prior work. In the context of shallow types, the pypy tracing JIT improved Reticulated [65] and the Graal partial-evaluation-based JIT improved Grace [16, 50]. In the context of concrete types, the HiggsCheck VM improved an implementation of SafeTypeScript [46, 47]. And in the context of deep/guarded/structural types, the Pycket tracing JIT greatly reduced the costs of higher-order checks in Typed Racket [1].

Static Python is one of many type systems for Python. Other optional type checkers include Pyre [37], mypy [57], PyType [18], and Pyright [38]. These checkers use types only for static analysis; optional types have no effect (positive or negative) at runtime. Another sound type system is Reticulated, which enforces structural types and pioneered the transient semantics [64, 66]. A rigorous comparison of Reticulated and Static Python is an important topic for future work. We conjecture that adapting code to Reticulated requires fewer edits than Static Python but delivers slower performance. Optimizing type systems for Python include Reticulated and mypyc [58]. The latter compiles Python source to C extension modules; it does not offer a pure-Python developer experience.

Although sound gradual types are rare in industry, there are at least three other languages that provide them. Dart 2 is a nominally-typed language with a dynamic type that is similar to a top type but statically allows all method calls [56]. C# has a dynamic type that delays type checking until runtime [3]. JS++ is typed JavaScript that allows untyped values using a catch-all external type [10]. None of these four implement gradual types as envisioned by researchers [19, 54, 62]; instead, they offer a pragmatic solution to interoperability problems.

## 8 Lessons

Static Python is an ambitious undertaking. Its developers are maintaining a fork of the Python 3.8 runtime, a type checker, an optimizing ahead-of-time compiler, and a method-based JIT compiler. The team has converted hundreds of untyped modules to Static Python and plans to convert thousands more in the future. And yet, Static Python is restricted to set attainable goals:

- The syntax of sound static types is relatively small compared to PEP 484. Higher-order types are absent. The focus is instead on first-order objects and standard data structures (Section 2.1).
- The type checker analyzes entire modules rather than individual statements, and although it supports a dynamic type, dynamic type-checked code does not have the same freedom as untyped code (Section 2.2).
- Shallow types accommodate Python code but offer coarse soundness guarantees. Concrete types are deeply sound, but add nonlocal constraints (Section 2.3).
- Gradual class hierarchies are permitted, but restricted to single inheritance. Static-to-static method overrides cannot reduce the precision of types (Section 2.4).
- Progressive primitive types are available to maximize performance (Section 2.5).

Many of these restrictions go against the grain of the mainstream research community, but they have been effective in practice. The current 3.7 % increase in Instagram web server CPU efficiency is a huge improvement and was obtained incrementally without any careful planning to avoid performance bottlenecks. By contrast, bottlenecks are a major concern for expressive, structurally-typed gradual languages [23, 55].

In conclusion, the gradual soundness approach of Static Python seems to be a promising way to realize the vision of gradual typing. Determining whether this conjecture holds more broadly calls for research on two fronts. First, Static Python must be applied to additional projects in a module-at-a-time manner. A systematic evaluation of a few small projects would be especially useful to find performance bottlenecks. Second, the Static Python approach should be adapted to other languages. An optionally-typed language such as TypeScript would be an ideal starting point.

**Data Availability Statement** The software that supports Section 4 of this paper is available via Zenodo [32] and Software Heritage [33].

**Acknowledgements** Static Python and the Cinder JIT were developed by Max Bernstein, John Biesnecker, Jacob Bower, Sinan Cepel, Tiansi Hu, Orvid King, Vladimir Matveev, William Meehan, Carl Meyer, Matt Page, Aniket Panse, Brett Simmers, An-drei Talaba, Dino Viehland, and Shiyu Wang. The Brown authors contributed to the language by modeling it independently; they are not part of the Static Python team.

Special thanks to Twitter, through which the two teams (specifically, Meyer and Krishnamurthi) met. Thanks also to Tobias Pape for going above and beyond to offer splendid assistance with the paper typesetting. This work was partly supported by Meta and by the US National Science Foundation. Greenman is supported by NSF grant 2030859 to the CRA for the CIFellows project.

## A GitHub Issues

The 26 GitHub issues that arose from our work may be found at the following three links. A list of issue numbers follows each link (all accessed 2022-05-17).

- https://github.com/facebookincubator/cinder/issues/created_by/LuKC1024 (N=20)
  #37, 39, 40, 41, 42, 43, 45, 46, 47, 49, 50, 51, 52, 53, 59, 60, 61, 62, 63, 65
- https://github.com/facebookincubator/cinder/issues/created_by/bennn (N=5)
  #35, 36, 54, 55, 71
- https://github.com/facebookincubator/cinder/issues/64 (N=1)
  #64

The critical soundness issues were: #36, 39, 53, 55, and 62.

## B On Migrations from Pyre to Static Python

Although the Instagram web server has extensive type annotations that are checked by Pyre, these annotations are not always enough to satisfy Static Python. In fact, code may have latent bugs that Pyre did not catch; we list common reasons for such bugs below:

- Pyre misses some bugs because it does not monitor run-time interactions with untyped code.
- Similarly, some uses of the Pyre dynamic type (`Any`) end up raising errors when Static Python monitors their interactions with precisely-typed code.
- Occurrences of the special comment `# pyre-fixme` disable Pyre checks.
- Some annotations intentionally lie to reduce the work of maintaining types. For example:
  - The standard typeshed repo of Python type annotations declares that a weak reference has the dynamic type [9]. This declaration is easy to use because it does not force programmers to handle the case where their weakly-held object has been collected, but it is also a potential source of bugs.
  - Mutable class attributes are unsoundly covariant in Pyre and other Python type checkers, including mypy [57].
  - The meta-type `Type` is unsoundly covariant as well in Pyre and others.

– Argument splatting, i.e., applying a function that expects positional arguments to a list (f(*lst)), is not statically checked either.

None of these issues are problems with Pyre. Quite naturally, it cannot find bugs in code that it does not analyze! The takeaway is simply that optional type systems can miss a variety of issues.

## C Skipped Static Python Regression Tests

As of January 9 (commit 6d61575), the Static Python test suite contains 802 tests. We use 265 of these tests for the model. The remaining 537 tests are skipped for the following reasons:

- 269 test primitive types.
- 62 test optional and keyword functions.
- 26 test the classloader.
- 24 test coroutines (async).
- 23 test list and dictionary comprehensions.
- 15 use format strings.
- 12 test class and object slots.
- The others ($N = 106$) test a long tail of other topics including byte strings, memory management, specific decorators (such as @staticmethod), and optimizations.

## D Fine-Grained Benchmark Data

The tables in this section present fine-grained data for the microbenchmark programs. Each benchmark has several versions of its code that accomplish the same work: the original untyped version (**Orig**) and a typed version (**T-Max**) converted for maximum performance under Static Python and the JIT. Some benchmarks have a basic typed version (**T-Min**) that adds only type annotations and necessary casts to the original code, without using any more advanced Static Python features such as primitive types or otherwise optimizing the code for Static Python. Fannkuch has a second basic version (**T-Min-2**) that uses just a few primitive integers where it can be done without significant changes to the code.

The other axes of our test matrix are **SP** (whether the Static Python compiler is used), **JIT** (whether the Cinder JIT is enabled), and **SF** (whether the JIT shadow-frame mode is enabled, which reduces Python frame allocation costs.) The matrix is not complete, since using the Static Python compiler on untyped code has no noticeable effect, and shadow-frame mode is only relevant under the JIT.

Benchmarks are run on a 64-bit CentOS Linux system, using a build of Cinder compiled with default configure settings. Overall process user time is measured via the time utility. Each benchmark/configuration is run 10 times (after an initial warmup run to update the bytecode cache file and thereby ignore compilation time), and all individual data points are recorded. All data points are in seconds.

JIT runs use a JIT list that includes only the benchmark code itself, to minimize the overhead of compiling standard library code not used in the benchmark.

- Benchmarks richards and fannkuch ran on commit d0d071
- deltablue ran on commit c9d14c
- nbody ran on commit 144a21

Fannkuch in particular is very numeric-heavy, so using unboxed primitive integers in the typed version can be a big win. For this to work maximally efficiently, we also need to avoid operations that don't yet support primitives and thus would require expensive boxing. In typed-and-optimized Fannkuch (**T-Max**), these changes move a lot of the work out of heavily-optimized C builtin methods (for list slice/insert/pop) and into simple Python code doing the equivalent operations manually with simple iteration and indexing. In the short run this is a massive pessimization for non-static Python (or even SP without the JIT, since currently we always use boxed integers in the interpreter loop.) In the long run it suggests that Static Python + JIT could make it feasible to implement many more of these core datastructure operations in pure Python rather than in C.

Because of this subtlety, we include benchmark results for both typed and untyped versions of each benchmark. In the short term, perhaps the most relevant comparison is the performance of Static Python on the typed benchmark vs the performance of nonstatic Python on the untyped benchmark; this gives a fair picture of expected perf gains converting code to Static Python with some willingness to optimize.

We also observe significant untapped opportunity to improve Static Python's performance; for example, 20 % of Typed Fannkuch SP+JIT time is spent on bounds-checking array accesses; most of this runtime bounds-checking cost could be eliminated if the Static Python compiler tracks known array sizes and integer values more thoroughly. And it would also be possible to support list slicing, list pop, and list insert with primitive arguments, thus avoiding the need to rewrite these operations in Python for maximum efficiency.

**T-Min** Fannkuch without the JIT is slow because of a known issue where we don't cache function lookups for INVOKE_FUNCTION in the interpreter loop, making the calls to list.insert and list.pop slower; this is fixable.

**Table 5** Richards microbenchmark data

| T-Max (SP JIT SF) | T-Max (SP JIT) | T-Max (SP) | T-Max (JIT SF) | T-Max (JIT) | T-Max () |
|---|---|---|---|---|---|
| 0.57 | 1.29 | 7.86 | 8.36 | 10.24 | 16 |
| 0.59 | 1.32 | 7.92 | 8.84 | 9.97 | 15.73 |
| 0.58 | 1.33 | 7.89 | 8.9 | 10.57 | 16.2 |
| 0.58 | 1.33 | 7.92 | 8.53 | 10.71 | 16.71 |
| 0.56 | 1.33 | 7.83 | 9.15 | 9.89 | 16.62 |
| 0.56 | 1.33 | 7.97 | 8.25 | 10.44 | 15.65 |
| 0.55 | 1.32 | 8.18 | 8.78 | 10.07 | 16.22 |
| 0.55 | 1.31 | 7.85 | 8.67 | 10.07 | 15.73 |
| 0.56 | 1.31 | 7.96 | 8.52 | 10.14 | 15.82 |
| 0.57 | 1.32 | 8 | 8.35 | 10.57 | 16.66 |

| T-Min (SP JIT SF) | T-Min (SP JIT) | T-Min (SP) | T-Min (JIT SF) | T-Min (JIT) | T-Min () |
|---|---|---|---|---|---|
| 1.41 | 2.68 | 7.67 | 5.58 | 7.97 | 13.32 |
| 1.38 | 2.6 | 7.9 | 5.92 | 7.25 | 13.59 |
| 1.39 | 2.64 | 7.74 | 5.54 | 7.64 | 13.83 |
| 1.38 | 2.65 | 7.8 | 6.06 | 7.24 | 13.95 |
| 1.37 | 2.62 | 7.68 | 5.52 | 7.2 | 13.5 |
| 1.39 | 2.58 | 7.8 | 5.63 | 7.27 | 13.52 |
| 1.36 | 2.65 | 7.79 | 5.74 | 7.17 | 13.34 |
| 1.43 | 2.66 | 7.82 | 5.56 | 7.54 | 13.68 |
| 1.53 | 2.63 | 7.9 | 5.59 | 6.89 | 13.79 |
| 1.37 | 2.71 | 7.95 | 5.74 | 6.91 | 13.5 |

| Orig (JIT SF) | Orig (JIT) | Orig () |
|---|---|---|
| 2.6 | 3.88 | 10.17 |
| 2.6 | 4.01 | 10.18 |
| 2.61 | 4.17 | 10.73 |
| 2.9 | 3.92 | 9.96 |
| 2.63 | 3.79 | 10.04 |
| 2.64 | 3.95 | 9.99 |
| 2.62 | 3.9 | 10.16 |
| 2.6 | 3.84 | 9.69 |
| 2.6 | 4.04 | 9.82 |
| 2.62 | 3.85 | 9.85 |

**Table 6** Fannkuch microbenchmark data

| T-Max (SP JIT SF) | T-Max (SP JIT) | T-Max (SP) | T-Max (JIT SF) | T-Max (JIT) | T-Max () |
|---|---|---|---|---|---|
| 1.27 | 1.39 | 34.69 | 40.74 | 41 | 48.23 |
| 1.32 | 1.25 | 34.35 | 40.23 | 40.94 | 49.03 |
| 1.28 | 1.26 | 34.67 | 40.74 | 40.65 | 49.42 |
| 1.23 | 1.31 | 35.71 | 40.94 | 39.7 | 49.35 |
| 1.28 | 1.38 | 34.85 | 39.59 | 40.46 | 49.18 |
| 1.26 | 1.34 | 33.96 | 40.45 | 39.99 | 47.79 |
| 1.26 | 1.23 | 34.65 | 40.78 | 40.97 | 49.7 |
| 1.27 | 1.35 | 33.83 | 40.96 | 41.78 | 47.92 |
| 1.23 | 1.4 | 36.23 | 40.49 | 40.29 | 48.68 |
| 1.23 | 1.27 | 36.05 | 40.72 | 40.11 | 48.29 |

| T-Min (SP JIT SF) | T-Min (SP JIT) | T-Min (SP) | T-Min (JIT SF) | T-Min (JIT) | T-Min () |
|---|---|---|---|---|---|
| 2.77 | 2.83 | 6.18 | 2.89 | 2.86 | 4.07 |
| 2.85 | 2.8 | 6.12 | 2.94 | 2.9 | 4.06 |
| 2.79 | 2.77 | 6.38 | 2.87 | 2.86 | 4.1 |
| 2.87 | 2.81 | 6.23 | 2.87 | 2.85 | 4.29 |
| 2.85 | 2.84 | 6.17 | 2.92 | 2.81 | 4.14 |
| 2.87 | 2.8 | 6.21 | 2.94 | 2.8 | 4.06 |
| 2.92 | 2.76 | 6.04 | 3.11 | 3.05 | 4.08 |
| 2.78 | 2.76 | 6.13 | 3.1 | 2.86 | 4.19 |
| 2.84 | 2.76 | 6.13 | 2.87 | 2.96 | 4.31 |
| 2.92 | 2.79 | 6.17 | 2.91 | 2.96 | 4.1 |

| T-Min-2 (SP JIT SF) | T-Min-2 (SP JIT) | T-Min-2 (SP) | T-Min-2 (JIT SF) | T-Min-2 (JIT) | T-Min-2 () |
|---|---|---|---|---|---|
| 2.48 | 2.52 | 7.28 | 3.16 | 3.08 | 4.59 |
| 2.52 | 2.48 | 7.28 | 3.26 | 3.08 | 4.51 |
| 2.47 | 2.53 | 7.24 | 3.11 | 3.18 | 4.59 |
| 2.48 | 2.57 | 7.35 | 3.18 | 3.14 | 4.73 |
| 2.5 | 2.56 | 7.34 | 3.14 | 3.12 | 4.63 |
| 2.43 | 2.47 | 7.17 | 3.23 | 3.09 | 4.53 |
| 2.47 | 2.69 | 7.18 | 3.41 | 3.21 | 4.48 |
| 2.47 | 2.45 | 7.14 | 3.43 | 3.14 | 4.53 |
| 2.43 | 2.55 | 7.23 | 3.08 | 3.24 | 4.53 |
| 2.52 | 2.49 | 7.2 | 3.16 | 3.18 | 4.45 |

| Orig (JIT SF) | Orig (JIT) | Orig () |
|---|---|---|
| 2.81 | 2.94 | 4.02 |
| 2.66 | 2.85 | 4.22 |
| 2.69 | 2.78 | 4.01 |
| 2.75 | 2.8 | 4.01 |
| 2.75 | 2.92 | 4.07 |
| 2.78 | 2.83 | 4.36 |
| 2.84 | 2.81 | 4.19 |
| 2.7 | 2.86 | 4.25 |
| 2.76 | 2.78 | 4.08 |
| 2.89 | 2.76 | 4.16 |

**Table 7** DeltaBlue microbenchmark data

| T-Max (SP JIT SF) | T-Max (SP JIT) | T-Max (SP) | T-Max (JIT SF) | T-Max (JIT) | T-Max () |
|---|---|---|---|---|---|
| 0.3 | 0.4 | 1.32 | 0.51 | 0.64 | 1.02 |
| 0.32 | 0.39 | 1.23 | 0.54 | 0.64 | 1 |
| 0.31 | 0.4 | 1.24 | 0.54 | 0.64 | 0.99 |
| 0.3 | 0.42 | 1.31 | 0.56 | 0.65 | 1.02 |
| 0.28 | 0.4 | 1.24 | 0.52 | 0.65 | 0.96 |
| 0.28 | 0.39 | 1.24 | 0.55 | 0.65 | 0.98 |
| 0.29 | 0.4 | 1.27 | 0.55 | 0.68 | 0.97 |
| 0.3 | 0.42 | 1.29 | 0.53 | 0.67 | 1 |
| 0.29 | 0.39 | 1.28 | 0.53 | 0.69 | 0.99 |
| 0.28 | 0.42 | 1.32 | 0.51 | 0.64 | 0.99 |

| T-Min (SP JIT SF) | T-Min (SP JIT) | T-Min (SP) | T-Min (JIT SF) | T-Min (JIT) | T-Min () |
|---|---|---|---|---|---|
| 0.6 | 0.74 | 1.32 | 0.69 | 0.82 | 1.27 |
| 0.59 | 0.74 | 1.32 | 0.67 | 0.84 | 1.27 |
| 0.57 | 0.71 | 1.31 | 0.7 | 0.89 | 1.23 |
| 0.58 | 0.74 | 1.23 | 0.73 | 0.82 | 1.26 |
| 0.6 | 0.71 | 1.32 | 0.71 | 0.84 | 1.26 |
| 0.59 | 0.73 | 1.4 | 0.71 | 0.84 | 1.23 |
| 0.56 | 0.7 | 1.37 | 0.71 | 0.85 | 1.35 |
| 0.58 | 0.69 | 1.3 | 0.71 | 0.82 | 1.31 |
| 0.58 | 0.69 | 1.27 | 0.69 | 0.81 | 1.34 |
| 0.6 | 0.73 | 1.39 | 0.68 | 0.8 | 1.33 |

| T-Min-2 (SP JIT SF) | T-Min-2 (SP JIT) | T-Min-2 (SP) | T-Min-2 (JIT SF) | T-Min-2 (JIT) | T-Min-2 () |
|---|---|---|---|---|---|
| 0.54 | 0.66 | 1.38 | 0.61 | 0.76 | 1.17 |
| 0.51 | 0.62 | 1.44 | 0.61 | 0.75 | 1.25 |
| 0.55 | 0.69 | 1.41 | 0.6 | 0.82 | 1.17 |
| 0.54 | 0.69 | 1.41 | 0.63 | 0.75 | 1.16 |
| 0.54 | 0.7 | 1.41 | 0.65 | 0.75 | 1.17 |
| 0.55 | 0.67 | 1.4 | 0.66 | 0.75 | 1.15 |
| 0.52 | 0.67 | 1.43 | 0.61 | 0.75 | 1.15 |
| 0.52 | 0.66 | 1.4 | 0.62 | 0.75 | 1.18 |
| 0.52 | 0.64 | 1.51 | 0.65 | 0.76 | 1.18 |
| 0.5 | 0.65 | 1.41 | 0.62 | 0.79 | 1.15 |

| Orig (JIT SF) | Orig (JIT) | Orig () |
|---|---|---|
| 0.64 | 0.78 | 1.22 |
| 0.64 | 0.81 | 1.25 |
| 0.63 | 0.81 | 1.23 |
| 0.62 | 0.79 | 1.2 |
| 0.67 | 0.81 | 1.23 |
| 0.7 | 0.8 | 1.23 |
| 0.68 | 0.78 | 1.34 |
| 0.66 | 0.79 | 1.22 |
| 0.69 | 0.85 | 1.23 |
| 0.68 | 0.79 | 1.24 |

**Table 8** Nbody microbenchmark data

| T-Max (SP JIT SF) | T-Max (SP JIT) | T-Max (SP) | T-Max (JIT SF) | T-Max (JIT) | T-Max () |
|---|---|---|---|---|---|
| 0.17 | 0.18 | 3.61 | 0.92 | 0.88 | 2.12 |
| 0.17 | 0.18 | 3.57 | 0.91 | 0.88 | 2.09 |
| 0.17 | 0.18 | 3.68 | 0.96 | 0.88 | 2.24 |
| 0.17 | 0.17 | 3.82 | 0.93 | 0.9 | 2.15 |
| 0.17 | 0.17 | 3.72 | 0.89 | 0.9 | 2.17 |
| 0.18 | 0.18 | 3.63 | 0.89 | 0.9 | 2.14 |
| 0.18 | 0.16 | 3.5 | 0.89 | 0.89 | 2.23 |
| 0.18 | 0.17 | 3.45 | 0.89 | 0.93 | 2.32 |
| 0.2 | 0.17 | 3.61 | 0.92 | 0.88 | 2.16 |
| 0.2 | 0.18 | 3.63 | 0.9 | 0.92 | 2.06 |

| T-Min (SP JIT SF) | T-Min (SP JIT) | T-Min (SP) | T-Min (JIT SF) | T-Min (JIT) | T-Min () |
|---|---|---|---|---|---|
| 0.8 | 0.81 | 1.3 | 0.83 | 0.74 | 1.2 |
| 0.8 | 0.8 | 1.28 | 0.78 | 0.74 | 1.22 |
| 0.81 | 0.8 | 1.31 | 0.75 | 0.77 | 1.22 |
| 0.87 | 0.86 | 1.29 | 0.74 | 0.76 | 1.18 |
| 0.85 | 0.8 | 1.26 | 0.76 | 0.77 | 1.19 |
| 0.81 | 0.81 | 1.3 | 0.77 | 0.75 | 1.22 |
| 0.86 | 0.79 | 1.25 | 0.76 | 0.75 | 1.17 |
| 0.87 | 0.83 | 1.24 | 0.74 | 0.77 | 1.16 |
| 0.81 | 0.81 | 1.3 | 0.74 | 0.74 | 1.17 |
| 0.82 | 0.83 | 1.23 | 0.76 | 0.76 | 1.16 |

| Orig (JIT SF) | Orig (JIT) | Orig () |
|---|---|---|
| 0.78 | 0.75 | 1.15 |
| 0.74 | 0.73 | 1.19 |
| 0.75 | 0.72 | 1.19 |
| 0.73 | 0.73 | 1.18 |
| 0.83 | 0.76 | 1.16 |
| 0.82 | 0.72 | 1.24 |
| 0.75 | 0.72 | 1.15 |
| 0.72 | 0.76 | 1.2 |
| 0.72 | 0.72 | 1.15 |
| 0.74 | 0.73 | 1.2 |

**Table 9** Exact commands to invoke each benchmark

| Benchmark | Command |
|---|---|
| Richards **T-Max** (**SP JIT SF**) | time ./python -X jit -X jit-list-file=Tools/benchmarks/jitlist_richards_static.txt -X jit-enable-jit-list-wildcards↩ -X jit-shadow-frame -X install-strict-loader Tools/benchmarks/richards_static.py 100 |
| Richards **T-Max** (**SP JIT**) | time ./python -X jit -X jit-list-file=Tools/benchmarks/jitlist_richards_static.txt -X jit-enable-jit-list-wildcards↩ -X install-strict-loader Tools/benchmarks/richards_static.py 100 |
| Richards **T-Max** (**SP**) | time ./python -X install-strict-loader Tools/benchmarks/richards_static.py 100 |
| Richards **T-Max** (**JIT SF**) | time ./python -X jit -X jit-list-file=Tools/benchmarks/jitlist_richards_static.txt -X jit-enable-jit-list-wildcards↩ -X jit-shadow-frame Tools/benchmarks/richards_static.py 100 |
| Richards **T-Max** (**JIT**) | time ./python -X jit -X jit-list-file=Tools/benchmarks/jitlist_richards_static.txt -X jit-enable-jit-list-wildcards↩ Tools/benchmarks/richards_static.py 100 |
| Richards **T-Max** () | time ./python Tools/benchmarks/richards_static.py 100 |
| Richards **T-Min** (**SP JIT SF**) | time ./python -X jit -X jit-list-file=Tools/benchmarks/jitlist_richards_static_basic.txt -X jit-enable-jit-list-wil↩ dcards -X jit-shadow-frame -X install-strict-loader Tools/benchmarks/richards_static_basic.py 100 |
| Richards **T-Min** (**SP JIT**) | time ./python -X jit -X jit-list-file=Tools/benchmarks/jitlist_richards_static_basic.txt -X jit-enable-jit-list-wil↩ dcards -X install-strict-loader Tools/benchmarks/richards_static_basic.py 100 |
| Richards **T-Min** (**SP**) | time ./python -X install-strict-loader Tools/benchmarks/richards_static_basic.py 100 |
| Richards **T-Min** (**JIT SF**) | time ./python -X jit -X jit-list-file=Tools/benchmarks/jitlist_richards_static_basic.txt -X jit-enable-jit-list-wil↩ dcards -X jit-shadow-frame Tools/benchmarks/richards_static_basic.py 100 |
| Richards **T-Min** (**JIT**) | time ./python -X jit -X jit-list-file=Tools/benchmarks/jitlist_richards_static_basic.txt -X jit-enable-jit-list-wil↩ dcards Tools/benchmarks/richards_static_basic.py 100 |
| Richards **T-Min** () | time ./python Tools/benchmarks/richards_static_basic.py 100 |
| Richards **Orig** (**JIT SF**) | time ./python -X jit -X jit-list-file=Tools/benchmarks/jitlist_main.txt -X jit-enable-jit-list-wildcards -X jit-sha↩ dow-frame Tools/benchmarks/richards.py 100 |
| Richards **Orig** (**JIT**) | time ./python -X jit -X jit-list-file=Tools/benchmarks/jitlist_main.txt -X jit-enable-jit-list-wildcards Tools/benc↩ hmarks/richards.py 100 |
| Richards **Orig** () | time ./python Tools/benchmarks/richards.py 100 |
| Fannkuch **T-Max** (**SP JIT SF**) | time ./python -X jit -X jit-list-file=Tools/benchmarks/jitlist_fannkuch_static.txt -X jit-enable-jit-list-wildcards↩ -X jit-shadow-frame -X install-strict-loader Tools/benchmarks/fannkuch_static.py 5 |
| Fannkuch **T-Max** (**SP JIT**) | time ./python -X jit -X jit-list-file=Tools/benchmarks/jitlist_fannkuch_static.txt -X jit-enable-jit-list-wildcards↩ -X install-strict-loader Tools/benchmarks/fannkuch_static.py 5 |
| Fannkuch **T-Max** (**SP**) | time ./python -X install-strict-loader Tools/benchmarks/fannkuch_static.py 5 |
| Fannkuch **T-Max** (**JIT SF**) | time ./python -X jit -X jit-list-file=Tools/benchmarks/jitlist_fannkuch_static.txt -X jit-enable-jit-list-wildcards↩ -X jit-shadow-frame Tools/benchmarks/fannkuch_static.py 5 |
| Fannkuch **T-Max** (**JIT**) | time ./python -X jit -X jit-list-file=Tools/benchmarks/jitlist_fannkuch_static.txt -X jit-enable-jit-list-wildcards↩ Tools/benchmarks/fannkuch_static.py 5 |
| Fannkuch **T-Max** () | time ./python Tools/benchmarks/fannkuch_static.py 5 |
| Fannkuch **T-Min** (**SP JIT SF**) | time ./python -X jit -X jit-list-file=Tools/benchmarks/jitlist_fannkuch_static_basic.txt -X jit-enable-jit-list-wil↩ dcards -X jit-shadow-frame -X install-strict-loader Tools/benchmarks/fannkuch_static_basic.py 5 |
| Fannkuch **T-Min** (**SP JIT**) | time ./python -X jit -X jit-list-file=Tools/benchmarks/jitlist_fannkuch_static_basic.txt -X jit-enable-jit-list-wil↩ dcards -X install-strict-loader Tools/benchmarks/fannkuch_static_basic.py 5 |
| Fannkuch **T-Min** (**SP**) | time ./python -X install-strict-loader Tools/benchmarks/fannkuch_static_basic.py 5 |
| Fannkuch **T-Min** (**JIT SF**) | time ./python -X jit -X jit-list-file=Tools/benchmarks/jitlist_fannkuch_static_basic.txt -X jit-enable-jit-list-wil↩ dcards -X jit-shadow-frame Tools/benchmarks/fannkuch_static_basic.py 5 |

**Table 9** Exact commands to invoke each benchmark *(continued)*

| Benchmark | Command |
|---|---|
| Fannkuch **T-Min** (**JIT**) | time ./python -X jit -X jit-list-file=Tools/benchmarks/jitlist_fannkuch_static_basic.txt -X jit-enable-jit-list-wildcards Tools/benchmarks/fannkuch_static_basic.py 5 |
| Fannkuch **T-Min** () | time ./python Tools/benchmarks/fannkuch_static_basic.py 5 |
| Fannkuch **T-Min** (**SP JIT SF**) | time ./python -X jit -X jit-list-file=Tools/benchmarks/jitlist_fannkuch_static_basic2.txt -X jit-enable-jit-list-wildcards -X jit-shadow-frame -X install-strict-loader Tools/benchmarks/fannkuch_static_basic2.py 5 |
| Fannkuch **T-Min** (**SP JIT**) | time ./python -X jit -X jit-list-file=Tools/benchmarks/jitlist_fannkuch_static_basic2.txt -X jit-enable-jit-list-wildcards -X install-strict-loader Tools/benchmarks/fannkuch_static_basic2.py 5 |
| Fannkuch **T-Min** (**SP**) | time ./python -X install-strict-loader Tools/benchmarks/fannkuch_static_basic2.py 5 |
| Fannkuch **T-Min** (**JIT SF**) | time ./python -X jit -X jit-list-file=Tools/benchmarks/jitlist_fannkuch_static_basic2.txt -X jit-enable-jit-list-wildcards -X jit-shadow-frame Tools/benchmarks/fannkuch_static_basic2.py 5 |
| Fannkuch **T-Min** (**JIT**) | time ./python -X jit -X jit-list-file=Tools/benchmarks/jitlist_fannkuch_static_basic2.txt -X jit-enable-jit-list-wildcards Tools/benchmarks/fannkuch_static_basic2.py 5 |
| Fannkuch **T-Min** () | time ./python Tools/benchmarks/fannkuch_static_basic2.py 5 |
| Fannkuch **Orig** (**JIT SF**) | time ./python -X jit -X jit-list-file=Tools/benchmarks/jitlist_main.txt -X jit-enable-jit-list-wildcards -X jit-shadow-frame Tools/benchmarks/fannkuch.py 5 |
| Fannkuch **Orig** (**JIT**) | time ./python -X jit -X jit-list-file=Tools/benchmarks/jitlist_main.txt -X jit-enable-jit-list-wildcards Tools/benchmarks/fannkuch.py 5 |
| Fannkuch **Orig** () | time ./python Tools/benchmarks/fannkuch.py 5 |
| DeltaBlue **T-Max** (**SP JIT SF**) | time ./python -X jit -X jit-list-file=Tools/benchmarks/jitlist_deltablue_static.txt -X jit-enable-jit-list-wildcards -X jit-shadow-frame -X install-strict-loader Tools/benchmarks/deltablue_static.py 100 |
| DeltaBlue **T-Max** (**SP JIT**) | time ./python -X jit -X jit-list-file=Tools/benchmarks/jitlist_deltablue_static.txt -X jit-enable-jit-list-wildcards -X install-strict-loader Tools/benchmarks/deltablue_static.py 100 |
| DeltaBlue **T-Max** (**SP**) | time ./python -X install-strict-loader Tools/benchmarks/deltablue_static.py 100 |
| DeltaBlue **T-Max** (**JIT SF**) | time ./python -X jit -X jit-list-file=Tools/benchmarks/jitlist_deltablue_static.txt -X jit-enable-jit-list-wildcards -X jit-shadow-frame Tools/benchmarks/deltablue_static.py 100 |
| DeltaBlue **T-Max** (**JIT**) | time ./python -X jit -X jit-list-file=Tools/benchmarks/jitlist_deltablue_static.txt -X jit-enable-jit-list-wildcards Tools/benchmarks/deltablue_static.py 100 |
| DeltaBlue **T-Max** () | time ./python Tools/benchmarks/deltablue_static.py 100 |
| DeltaBlue **T-Min** (**SP JIT SF**) | time ./python -X jit -X jit-list-file=Tools/benchmarks/jitlist_deltablue_static_basic.txt -X jit-enable-jit-list-wildcards -X jit-shadow-frame -X install-strict-loader Tools/benchmarks/deltablue_static_basic.py 100 |
| DeltaBlue **T-Min** (**SP JIT**) | time ./python -X jit -X jit-list-file=Tools/benchmarks/jitlist_deltablue_static_basic.txt -X jit-enable-jit-list-wildcards -X install-strict-loader Tools/benchmarks/deltablue_static_basic.py 100 |
| DeltaBlue **T-Min** (**SP**) | time ./python -X install-strict-loader Tools/benchmarks/deltablue_static_basic.py 100 |
| DeltaBlue **T-Min** (**JIT SF**) | time ./python -X jit -X jit-list-file=Tools/benchmarks/jitlist_deltablue_static_basic.txt -X jit-enable-jit-list-wildcards -X jit-shadow-frame Tools/benchmarks/deltablue_static_basic.py 100 |
| DeltaBlue **T-Min** (**JIT**) | time ./python -X jit -X jit-list-file=Tools/benchmarks/jitlist_deltablue_static_basic.txt -X jit-enable-jit-list-wildcards Tools/benchmarks/deltablue_static_basic.py 100 |
| DeltaBlue **T-Min** () | time ./python Tools/benchmarks/deltablue_static_basic.py 100 |
| DeltaBlue **T-Min** (**SP JIT SF**) | time ./python -X jit -X jit-list-file=Tools/benchmarks/jitlist_deltablue_static_basic2.txt -X jit-enable-jit-list-wildcards -X jit-shadow-frame -X install-strict-loader Tools/benchmarks/deltablue_static_basic2.py 100 |
| DeltaBlue **T-Min** (**SP JIT**) | time ./python -X jit -X jit-list-file=Tools/benchmarks/jitlist_deltablue_static_basic2.txt -X jit-enable-jit-list-wildcards -X install-strict-loader Tools/benchmarks/deltablue_static_basic2.py 100 |
| DeltaBlue **T-Min** (**SP**) | time ./python -X install-strict-loader Tools/benchmarks/deltablue_static_basic2.py 100 |

**Table 9** Exact commands to invoke each benchmark *(continued)*

| Benchmark | Command |
|---|---|
| DeltaBlue **T-Min** (**JIT SF**) | time ./python -X jit -X jit-list-file=Tools/benchmarks/jitlist_deltablue_static_basic2.txt -X jit-enable-jit-list-wildcards -X jit-shadow-frame Tools/benchmarks/deltablue_static_basic2.py 100 |
| DeltaBlue **T-Min** (**JIT**) | time ./python -X jit -X jit-list-file=Tools/benchmarks/jitlist_deltablue_static_basic2.txt -X jit-enable-jit-list-wildcards Tools/benchmarks/deltablue_static_basic2.py 100 |
| DeltaBlue **T-Min** () | time ./python Tools/benchmarks/deltablue_static_basic2.py 100 |
| DeltaBlue **Orig** (**JIT SF**) | time ./python -X jit -X jit-list-file=Tools/benchmarks/jitlist_main.txt -X jit-enable-jit-list-wildcards -X jit-shadow-frame Tools/benchmarks/deltablue.py 100 |
| DeltaBlue **Orig** (**JIT**) | time ./python -X jit -X jit-list-file=Tools/benchmarks/jitlist_main.txt -X jit-enable-jit-list-wildcards Tools/benchmarks/deltablue.py 100 |
| DeltaBlue **Orig** () | time ./python Tools/benchmarks/deltablue.py 100 |
| NBody **T-Max** (**SP JIT SF**) | time ./python -X jit -X jit-list-file=Tools/benchmarks/jitlist_nbody_static.txt -X jit-enable-jit-list-wildcards -X jit-shadow-frame -X install-strict-loader Tools/benchmarks/nbody_static.py |
| NBody **T-Max** (**SP JIT**) | time ./python -X jit -X jit-list-file=Tools/benchmarks/jitlist_nbody_static.txt -X jit-enable-jit-list-wildcards -X install-strict-loader Tools/benchmarks/nbody_static.py |
| NBody **T-Max** (**SP**) | time ./python -X install-strict-loader Tools/benchmarks/nbody_static.py |
| NBody **T-Max** (**JIT SF**) | time ./python -X jit -X jit-list-file=Tools/benchmarks/jitlist_nbody_static.txt -X jit-enable-jit-list-wildcards -X jit-shadow-frame Tools/benchmarks/nbody_static.py |
| NBody **T-Max** (**JIT**) | time ./python -X jit -X jit-list-file=Tools/benchmarks/jitlist_nbody_static.txt -X jit-enable-jit-list-wildcards Tools/benchmarks/nbody_static.py |
| NBody **T-Max** () | time ./python Tools/benchmarks/nbody_static.py |
| NBody **T-Min** (**SP JIT SF**) | time ./python -X jit -X jit-list-file=Tools/benchmarks/jitlist_nbody_static_basic.txt -X jit-enable-jit-list-wildcards -X jit-shadow-frame -X install-strict-loader Tools/benchmarks/nbody_static_basic.py |
| NBody **T-Min** (**SP JIT**) | time ./python -X jit -X jit-list-file=Tools/benchmarks/jitlist_nbody_static_basic.txt -X jit-enable-jit-list-wildcards -X install-strict-loader Tools/benchmarks/nbody_static_basic.py |
| NBody **T-Min** (**SP**) | time ./python -X install-strict-loader Tools/benchmarks/nbody_static.py |
| NBody **T-Min** (**JIT SF**) | time ./python -X jit -X jit-list-file=Tools/benchmarks/jitlist_nbody_static_basic.txt -X jit-enable-jit-list-wildcards -X jit-shadow-frame Tools/benchmarks/nbody_static_basic.py |
| NBody **T-Min** (**JIT**) | time ./python -X jit -X jit-list-file=Tools/benchmarks/jitlist_nbody_static.txt -X jit-enable-jit-list-wildcards Tools/benchmarks/nbody_static.py |
| NBody **T-Min** () | time ./python Tools/benchmarks/nbody_static_basic.py |
| NBody **Orig** (**JIT SF**) | time ./python -X jit -X jit-list-file=Tools/benchmarks/jitlist_main.txt -X jit-enable-jit-list-wildcards -X jit-shadow-frame Tools/benchmarks/nbody.py |
| NBody **Orig** (**JIT**) | time ./python -X jit -X jit-list-file=Tools/benchmarks/jitlist_main.txt -X jit-enable-jit-list-wildcards Tools/benchmarks/nbody.py |
| NBody **Orig** () | time ./python Tools/benchmarks/nbody.py |

## About the authors

**Kuang-Chen Lu** (LuKuangchen1024@gmail.com) is a PhD student at Brown University.

**Ben Greenman** (benjamin.l.greenman@gmail.com) is a postdoc at Brown University.

**Carl Meyer** (carljm@fb.com) is a software engineer on the Python Language & Runtime team at Meta.

**Dino Viehland** (dinoviehland@fb.com) is a software engineer at Meta working on the Cinder Python runtime. Previously he has worked on other Python related efforts including Python Tools for Visual Studio and IronPython.

**Aniket Panse** (aniketpanse@fb.com) is a software engineer on the Python Language & Runtime team at Meta.

**Shriram Krishnamurthi** (shriram@brown.edu) is the Vice President of Programming Languages (no, not really) at Brown University.